\documentclass[conference]{IEEEtran}
\IEEEoverridecommandlockouts

\usepackage{cite}
\usepackage{amsmath,amssymb,amsfonts}
\usepackage{verbatim}
\usepackage{stmaryrd}
\usepackage{algorithmic}
\usepackage{graphicx}
\usepackage{textcomp}
\usepackage{lipsum}
\usepackage{xcolor}
\usepackage{bbold}
\usepackage{dsfont}
\def\BibTeX{{\rm B\kern-.05em{\sc i\kern-.025em b}\kern-.08em
    T\kern-.1667em\lower.7ex\hbox{E}\kern-.125emX}}

\newtheorem{theo}{Theorem}
\newtheorem{prop}{Proposition}

\newtheorem{lemm}{Lemma}
\newtheorem{defi}{Definition}

 \newtheorem{remark}{Remark}  
 \allowdisplaybreaks
\begin{document}

\newcommand{\Fonction}[5]{
\begin{array}{l|ccl}
#1 : & #2 & \longrightarrow & #3 \\
    & #4 & \longmapsto & #5
\end{array}}

\newcommand{\norme}[1]{\left\Vert #1\right\Vert}
\newcommand{\vertiii}[1]{{\left\vert\kern-0.25ex\left\vert\kern-0.25ex\left\vert #1 
    \right\vert\kern-0.25ex\right\vert\kern-0.25ex\right\vert}}

\renewcommand\l[2]{\xrightarrow[#1 \to #2]{}}
\newcommand\eps{\varepsilon}
\newcommand\exist{\exists\,}
\newcommand{\dint}{\text{d}}
\renewcommand\emptyset{\varnothing}

\newcommand\NN{\textbf{N}} 
\newcommand\ZZ{\textbf{Z}} 
\newcommand\QQ{\textbf{Q}} 
\newcommand\RR{\textbf{R}} 
\newcommand\CC{\textbf{C}} 
\newcommand\PP{\textbf{P}}
\newcommand\KK{\textbf{K}} 
\newcommand\FF{\textbf{F}}
\newcommand\EE{\textbf{E}}
\newcommand\LL{\textbf{L}}
\newcommand\VV{\textbf{V}}
\newcommand\ZZZ[1]{\textbf{Z}/#1\textbf{Z}}
\newcommand\FC{\mathcal{C}}
\newcommand\FCPI{\mathcal{C}_{2\pi}}
\newcommand\MM[2]{M_{#1}(#2)}
\newcommand\MGL[2]{GL_{#1}(#2)}
\newcommand\MO[2]{O_{#1}(#2)}
\newcommand\MS[2]{S_{#1}(#2)}
\newcommand\MSP[2]{S_{#1}^+(#2)}
\newcommand\MSPP[2]{S_{#1}^{++}(#2)}
\newcommand\MH[1]{H_{#1}(\CC)}
\newcommand*{\QEDB}{\null\nobreak\hfill\ensuremath{\square}}
\newcommand\suite[2]{(#1_{#2})_{#2\in\NN}}
\newcommand\CM{\mathcal{M}}
\newcommand\CQ{\mathcal{Q}}
\newcommand\CT{\mathcal{T}}
\newcommand\excur{\mathbbm{e}}
\newcommand\Mcarte{\mathfrak{m}}
\newcommand\Qcarte{\mathfrak{q}}
\newcommand\floor[1]{\lfloor {#1}\rfloor}
\newcommand\ceil[1]{\lceil {#1}\rceil}
\newcommand{\Lim}[1]{\raisebox{0.5ex}{\scalebox{0.8}{$\displaystyle \lim_{#1}\;$}}}
\newcommand{\Limsup}[1]{\raisebox{0.5ex}{\scalebox{0.8}{$\displaystyle \limsup_{#1}\;$}}}
\newcommand{\Liminf}[1]{\raisebox{0.5ex}{\scalebox{0.8}{$\displaystyle \liminf_{#1}\;$}}}
\newcommand\indic{\mathbb{1}}
\newcommand{\mw}[1]{{\color{blue}#1}}

\title{Typicality with Feedback}

\author{\IEEEauthorblockN{Thomas Sturma, Michèle Wigger}
\IEEEauthorblockA{\textit{LTCI}, 
\textit{Télécom Paris}, 
Palaiseau, France 
}}

\maketitle

\begin{abstract}
   The main objective of this paper is to analyze a closed-loop  feedback system where a transmitter probes  a discrete memoryless channel (DMC) and can adapt its inputs based on the previous channel outputs. We prove that, regardless of the transmitter’s strategy, the conditional type of the outputs given the inputs remains close to the DMC transition law \(P_{Y|X}\). This general result enables the study of fundamental limits in certain adaptive systems.

    As an application, we establish a  converse result for an integrated sensing and communication (ISAC) model. In this setting, the transmitter also functions as a radar receiver, aiming to simultaneously transmit a message over the channel and estimate the channel state from the backscattered feedback signals. We show that the fundamental limits of the closed loop system are the same as of the open-loop system where the transmitter can use the feedback signal to estimate the state but not to produce adaptive channel inputs. This result holds as long as the sum of the admissible-average-decoding-error-probability, denoted \(\epsilon\), and the admissible-excess-distortion-probability,  denoted \(\delta\), is below \(1\), i.e., \(\delta +\epsilon<1\). 
\end{abstract}


\section{Introduction}

In this paper, we investigate a closed-loop  system where a transmitter probes  a discrete memoryless channel (DMC) and the transmitter can adapt its inputs based on past channel outputs, see Figure~\ref{fig:channel_probing}. We show that, irrespective 
 of  the transmitter's strategy in producing the inputs, the empirical conditional distribution---or conditional type---of the channel outputs given the inputs remains close to the underlying channel law \( P_{Y|X} \). For open-loop systems, where  channel inputs cannot depend on past outputs, the desired statement follows  from Chebychef's inequality in a straightforward way and in the asymptotic limit is nothing but the weak law of large numbers. With closed-loop strategies, the situation is more complicated because the transmitter can adapt its inputs so as to maximize the deviation between the empirical distribution and the true channel law $P_{Y|X}$. Proving the desired statement on the conditional type  requires a different proof. 
 
Specifically,  we identify for each input-output pair $(a,b)$ the transmitter strategy that maximizes the probability that the conditional type of output $b$ given input $a$ deviates from \( P_{Y|X}(b|a) \) by more than a fixed threshold \( \mu \).  The optimal strategy has a simple form: the transmitter feeds input $a$ to the channel until the number of observed $b$ symbols deviates from $b$ by more than $n$ times the threshold $\mu$, at which point it starts feeding non-``$a$" symbols to the channel until the number of channel uses $n$ is reached.  Optimality of this strategy is proved by first noting the bijection between input strategies and labels on a full $|\mathcal{Y}|$-ary tree of depth $n$, and then deriving the labeling on the tree that maximizes the desired probability. We finally analyze this probability for the described strategy using a simple inequality on sums of i.i.d. random variables. 

We subsequently use the established result on the conditional types to prove a converse result for  an Integrated Sensing and Communication (ISAC) system where the transmitter sends a message over a state-dependent discrete memoryless channel (SDMC) with feedback and from the feedback signals also reconstructs the state sequence up to desired distortion. Our converse proof is based on change of measure arguments \cite{HanKobayashi,b1,b2,b3,watanabe,b7} and establishes that for any permissible decoding error probability $\epsilon$  and excess distortion probability $\delta$ whose sum is below $1$, i.e., $\epsilon+\delta<1$, the set of data rates and reconstruction distortions that are simultaneously achievable, coincides with this set in an open-loop system where the transmitter  is not allowed to adapt its inputs in function of the feedback signals. Our results assume perfect feedback, and extend to noisy or degraded versions thereof. Extension to arbitrary generalized feedback signals as considered in \cite{kobayashi2018joint,b6,b7} is left for future work.

\textit{Notation.} 
Sets are denoted using calligraphic fronts such as $\mathcal{X}$ and $\mathcal{Y}$. All random variables are assumed finite. For $i\leq j$ two positive integers, sequences of random variables $(X_i,\dots , X_j)$ and realizations $(x_i,\dots , x_j)$ are abbreviated by $X_i^j$ and $x_i^j$ respectively. If $i=1$, we also use $X^j$ and $x^j$ instead of $X_i^j$ and $x_i^j$. The probability of an event $A$ and the expectation of a random variable $X$ are denoted by $\PP(A)$ and $\EE[X]$ respectively. Moreover, throughout this paper, $2^{nR}$ will denote the integer $\lfloor 2^{nR} \rfloor$. We denote by $\pi_{x^ny^n}$ the joint type of $(x^n,y^n)$ defined in \cite{Csiszarbook}. The marginal type of $x^n$ is denoted  $\pi_{x^n}$. Finally, we will use the Landau notation $o(1)$ to indicate any function that tends to 0 when $n\rightarrow +\infty$.



\section{Typicality Through a Channel  with Feedback}
\begin{figure}[h!]
    \centering
    \includegraphics[height=1.9cm]{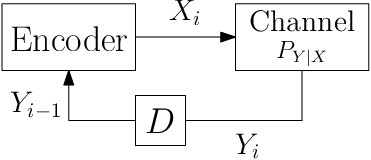}
    \caption{Typicality through a Channel with Feedback}
    \label{fig:channel_probing}
\end{figure}
 
Consider the setup in Figure~\ref{fig:channel_probing}. An encoder called `Alex' produces inputs $X_1,\ldots, X_n$ to  a channel $P_{Y|X}$ with feedback. That is, it can observe the past outputs of the channel and choose the inputs in function thereof: \begin{equation} 
X_i=h_i( Y_1, \ldots, Y_{i-1}), \quad i\in\{1,\ldots, n\},
\end{equation}
for some functions $h_i(\cdot)$ on appropriate domains. 
The following lemma asserts that whatever strategy, i.e., sequence of functions $\{h_1, \ldots, h_n\}$, Alex chooses, the conditional type of the outputs $Y^n$ given the inputs $X^n$ is close to the channel law $P_{Y|X}$ with high probability. 
\begin{lemm}
    \label{lemmimp}
Irrespective of the choice of the functions $\{h_1, \ldots, h_n\}$,     for any $(a,b)\in\mathcal{X}\times\mathcal{Y}$ and small value $\mu>0$:
    \begin{multline}
        \PP\bigl(\left|\pi_{X^nY^n}(a,b)-\pi_{X^n}(a)P_{Y|X}(b)\right|>\mu\bigr)\leq\dfrac{1}{4n\mu^2}.
    \end{multline}
\end{lemm}
While without feedback this statement  holds by Chebychef's inequality, the proof is slightly trickier with feedback. 

\subsection{Proof of Lemma~\ref{lemmimp}}
 
Define for any strategy  
$\mathbf{h}=(h_1(\cdot), \ldots, h_n(\cdot))$ and any   $(a,b)\in\mathcal{X}\times\mathcal{Y}$, the random score
\begin{IEEEeqnarray}{rCl}
        S_n(\mathbf{h},a,b)= \sum_{k=1}^n\indic\{X_k=a\}(\indic\{Y_k=b\}-P_{Y|X}(b|a)) , \IEEEeqnarraynumspace
\end{IEEEeqnarray}
for $\{(X_k,Y_k)\}_{k=1}^n$ generated according to the strategy $\mathbf{h}$ and the channel $P_{Y|X}$.

Notice the equivalence: 
\begin{IEEEeqnarray}{rCl}
\lefteqn{ \PP\bigl(\left|\pi_{X^nY^n}(a,b)-\pi_{X^n}(a)P_{Y|X}(b|a)\right|>\mu\bigr)} \nonumber\\
&=&  \PP\left(\left|\sum_{i=1}^n \indic\{X_i=a\}\left(\indic\{Y_i=b\}-P_{Y|X}(b|a)\right)\right|>n\mu\right)\nonumber\\
& = &\PP\left( |S_n(\mathbf{h},a,b) |>n \mu\right).   \end{IEEEeqnarray}
We shall prove that for any   $(a,b)\in\mathcal{X}\times\mathcal{Y}$ 
\begin{equation}\label{eq:max}
\max_{\mathbf{h}}  \PP\left(| S_n(\mathbf{h},a,b)| >n \mu\right) \leq \frac{1}{4 n \mu^2},
\end{equation} 
implying the result in Lemma~\ref{lemmimp}. To this end, we first identify the maximizing strategy $\mathbf{h}$  for any given pair $(a,b)$, and then determine the probability that its score function exceeds $n \mu$. 

\begin{remark}
    Following a reviewer's suggestion, we can prove~\eqref{eq:max} by noting that $\{S_n(\mathbf{h},a,b)\}_n$ forms a martingale and then applying Doob's inequality. However, the following proof is stronger, as it identifies the optimal strategy.
\end{remark}

\subsubsection{A maximizing strategy $\mathbf{h}$}

Fix   $(a,b)\in\mathcal{X}\times\mathcal{Y}$.
For a given strategy $\mathbf{h}$, define for each 
$i\in\{1,\ldots, n\}$ and each tuple $y^{i}$:
    \begin{equation}
        s_i(y^{i},\mathbf{h})=\sum_{\ell=1}^{i}\indic\{x_\ell=a\}(\indic\{y_\ell=b\}-P_{Y|X}(b|a)),
    \end{equation}
    where $x_{\ell}=h_\ell(y^{\ell-1})$.

    \begin{lemm}\label{lem:optimal}
The following strategy is a maximizer of \eqref{eq:max}:   
  \begin{IEEEeqnarray}{rCl}h^*_k(y_1,\ldots, y_{k-1})=\begin{cases} 
            a, & \text{if }|s_{k-1}(y^{k-1}, \mathbf{h})|< n\mu\\
         \neq a & \text{otherwise},        \end{cases} \label{eq:optimal}
    \end{IEEEeqnarray}
    where $\neq a$ indicates  any symbol other than $a$. 
\end{lemm}
\begin{IEEEproof}[Proof of Lemma~\ref{lem:optimal}]
We start by
 noticing that each strategy  $\mathbf{h}$ can be represented as  labels on all nodes (including the vertex node but excluding leaves) of a  full $|\mathcal{Y}|$-ary rooted tree of depth $n$ where the $|\mathcal{Y}|$ edges below a node are labeled by the $|\mathcal{Y}|$ distinct elements of $\mathcal{Y}$. We simply use label  $h_k(y_1,\ldots, y_{k-1})$   for the level-$k$ node $v$ that is reached from the root through the edges labeled $y_1,\ldots, y_{k-1}$.  Figure~\ref{Arbre1} illustrates for example the tree (for $n=2$ and  $\mathcal{Y}=\{0,1,2\}$) and the node labels  corresponding to  strategy
 \begin{IEEEeqnarray}{rCl}
 h_1(\emptyset)&=&1\\
 h_2(y_1)&=&\begin{cases} 0 & \textnormal{ if } y_1 \in\{0,2\} \\
 1 & \textnormal{ if } y_1=1.
 \end{cases}
 \end{IEEEeqnarray} 
 
\begin{figure}
    \centering
    \includegraphics[height=2.3cm]{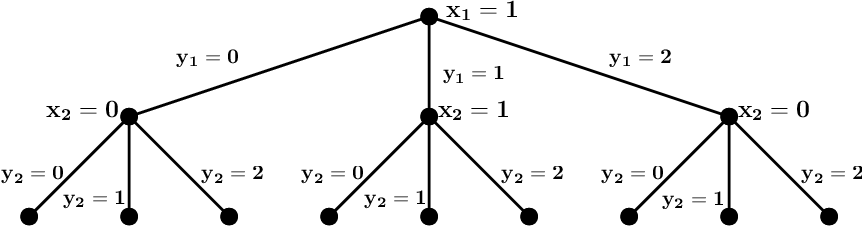}
s    \caption{Example of a tree for $n=2$}
    \label{Arbre1}
\end{figure}

%
A second example (for $n=3$, $|\mathcal{X}|=2$ and $|\mathcal{Y}|=3$)  is provided in Figure~\ref{Arbre4}, where for readability, we do not show the labels on the edges; they always correspond to the  elements of $\mathcal{Y}$ in the same order. 
    \begin{figure}
        \centering
        \includegraphics[height=3.2cm]{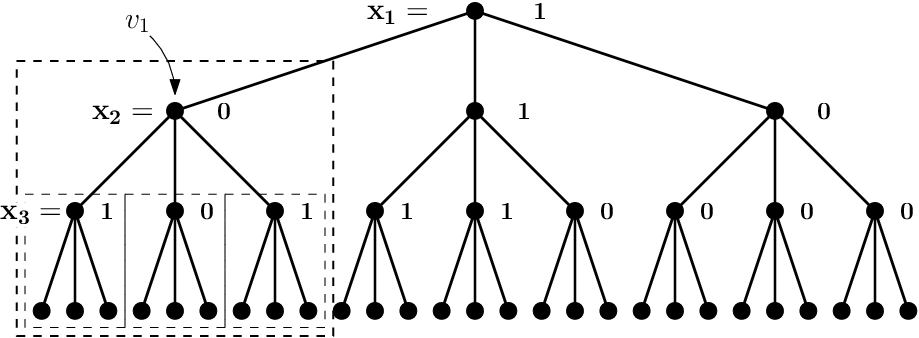}
        \caption{Example of a labeled tree for $n=3$, $|\mathcal{X}|=2$ and $|\mathcal{Y}|=3$.}
        \label{Arbre4}
    \end{figure}


%
%
We shall refer  by $\mathcal{T}_{\mathbf{h}}$ to the tree corresponding to a strategy $\mathbf{h}$. Notice the following facts for this tree $\mathcal{T}_{\mathbf{h}}$: 
\begin{itemize} 
\item  There is a one-to-one correspondance between the leaves of the tree and the  output sequences $y^n$.  We will thus simply refer to the sequences $y^n$ as the leaves of the tree. 
\item The 
 score $s_n(y^n, \mathbf{h})$ of a given leave node is obtained by running from the root node of the tree $\mathcal{T}_{\mathbf{h}}$ to leave $y^n$, and  by summing up for each edge  the term $(\indic\{y_k=b\}-P_{Y|X}(b|a))$ if its parent node  is $a$ and 0  otherwise. We will thus also write $s_i(y^i, \mathcal{T}_{\mathbf{h}})$ instead of  $s_i(y^i, {\mathbf{h}})$. 
 
 For a general labeled tree $\mathcal{T}$, we define 
 \begin{IEEEeqnarray}{rCl}
  \lefteqn{      s_i(y^{i},\mathcal{T})}\nonumber \\
  &\triangleq &\sum_{\ell=1}^{i}\indic\{x_\ell(y^{\ell-1})=a\}(\indic\{y_\ell=b\}-P_{Y|X}(b|a)), \IEEEeqnarraynumspace
\end{IEEEeqnarray}
    where $x_{\ell}(y^{\ell-1})$ is the label in $\mathcal{T}$ at the node on the path from the root indicated by $y_1, \ldots, y_{\ell-1}$.
 
 \item The success probability  $\PP(|S_n(\mathbf{h},a,b)| > n \mu)$ of strategy $\mathbf{h}$ is  directly  obtained from the tree $\mathcal{T}_{\mathbf{h}}$ by summing over the probabilities of the leaves with sufficiently high scores.
 
 We generally define for any labeled tree $\mathcal{T}$: 
 \begin{equation} \mathcal{S}(\mathcal{T}) \triangleq \{y^n \colon |s_n(y^n, \mathcal{T})|> n \mu\}, 
 \end{equation} 
 and the success probability of the tree
 \begin{IEEEeqnarray}{rCl} \label{eq:PT}
 P(\mathcal{T})
&\triangleq &  \sum_{y^n \in \mathcal{S}(\mathcal{T})}\;\; \prod_{i=1}^nP_{Y|X}(y_i| x_i(y^{i-1})), \IEEEeqnarraynumspace
\end{IEEEeqnarray}
where $x_i(y^{i-1})$ is the label on $\mathcal{T}$ of the node on the path from the root indicated by $y_1, \ldots, y_{i-1}$. Then, 
\begin{equation} 
P(\mathcal{T}_{\mathbf{h}})=\PP(|S_n(\mathbf{h},a,b)| > n \mu).
\end{equation} 
\end{itemize} 
Given these observations, we will denote by $\mathcal{F}$ the set of all labeled trees, and the bijection between the elements of  $\mathcal{F}$ and the set of  strategies. 
 
After these preliminary remarks, we tackle the proof of Lemma~\ref{lem:optimal}.  Notice that it is equivalent to proving that there exists an optimal labeled tree $\mathcal{T}^* \in \mathcal{F}$, i.e.,  $P(\mathcal{T}^*)= \max_{\mathcal{T}\in \mathcal{F}} P(\mathcal{T})$, with the following properties: 
 \begin{enumerate} 
 \item The tree  $\mathcal{T}^*$ is \emph{well-ordered}, meaning only nodes labeled $a$ can have descendants labeled $a$.  
  \item A level-$k$ node of $\mathcal{T}^*$ is labeled $a$ if, and  only if, $|s_{k}(y^{k}, \mathcal{T}^*)|\leq n \mu$,   where $y^{k}$ indicates the labels of the path from the root to this node.
 \end{enumerate} 

To show that 1.) holds, we start with any optimal tree $\mathcal{T}_0\in\mathcal{F}$ and iteratively construct new trees $\mathcal{T}_1, \mathcal{T}_2, \ldots$  so that $P(\mathcal{T}_k) \geq P(\mathcal{T}_{k-1})$ for any $k\geq 1$.  Moreover,  after a  finite number of steps $\kappa$ the tree $\mathcal{T}_{\kappa}$ is well-ordered and optimal.

Tree $\mathcal{T}_k$ is obtained from $\mathcal{T}_{k-1}$ as follows: 
\begin{itemize} 
\item Pick a vertex $v_k$ from tree $\mathcal{T}_{k-1}$ that is labeled $\tilde{a}\neq a$ but for which all parents and one of these children are labeled $a$.  (For example, consider vertex $v_1$ in Figure~\ref{Arbre4}.)
\item For any $y\in\mathcal{Y}$, denote by   $\mathcal{B}_y$ the subtree starting at the child of $v_k$ along the edge $y$. Label all  leave nodes of $\mathcal{B}_y$ by  $\tilde{a}$  and  augment all of these leaves with $|\mathcal{Y}|$ new children, where the edges  are labeled by all elements of $\mathcal{Y}$. ( Figure~\ref{Arbre4} indicates the trees $\mathcal{B}_0, \mathcal{B}_1, \mathcal{B}_2$ for node $v_1$ with dashed boxes and Figure~\ref{fig:augmented_tree} depicts the augmented tree of $\mathcal{B}_0$.) 
\item Starting from $\mathcal{T}_{k-1}$, create a new \emph{random} tree $\tilde{\mathcal{A}}$ as follows. For each $y\in\mathcal{Y}$, with probability $P_{Y|X}(y|\tilde{a})$ replace  node $v_k$ and its underlying subtree by the augmented tree  of $\mathcal{B}_y$. (Figure~\ref{fig:realization_tree} depicts the realization of $\tilde{\mathcal{A}}$ constructed with the augmented of $\mathcal{B}_0$.)
\item Among all $|\mathcal{Y}|$ possible realizations of the tree $\tilde{\mathcal{A}}$, let $\mathcal{T}_k$ be the one that has largest probability of success $P(\cdot)$.
 \end{itemize} 

\begin{figure}
    \centering
    \includegraphics[scale=0.7]{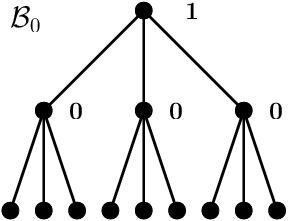}
    \caption{The augmented tree of $\mathcal{B}_0$}
    \label{fig:augmented_tree}
\end{figure}

\begin{figure}
    \centering
    \includegraphics[scale=0.57]{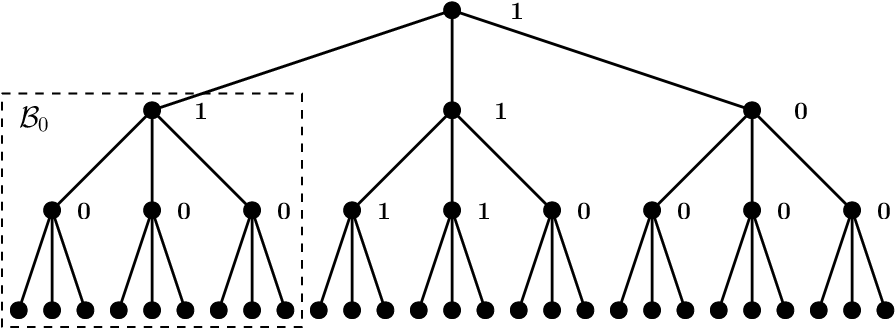}
    \caption{The realization of $\tilde{\mathcal{A}}$ constructed with the augmented tree of $\mathcal{B}_0$}
    \label{fig:realization_tree}
\end{figure}

It can be verified that  for each iteration $k$:
\begin{equation} \label{eq:ineq} 
P(\mathcal{T}_{k-1})= P(\tilde{\mathcal{A}})\leq P(\mathcal{T}_{k}).
\end{equation} 
Above inequality holds simply because the best is no worse than the average. The equality is proved in Appendix~\ref{app:same}

.

%
%
To prove 2), notice that once 1) is established, we can focus on 
 well-ordered trees only. By the property of these trees, the path from the root to any leaf $\bar{y}^n$ starts with a certain number of $a$-labels on the nodes, and then once it reaches   a specific node $v^*(\bar{y}^n)$ it switches to   non-$a$ labels for $v^*(\bar{y}^n)$ and all its descendants. Notice that  in an ordered tree $\mathcal{T}_{\textnormal{ord}}$,  all leaves $y^n$ that are descendants of a switching node $v^*(\bar{y}^n)$ have same  score as $\bar{y}^n$: $s_n(y^n,\mathcal{T}_{\textnormal{ord}})= s_n(\bar{y}^n,\mathcal{T}_{\textnormal{ord}})$. 
 
Fix an ordered tree $\mathcal{T}$ and a leave $\bar{y}^n$,  and let $k$ be the depth of ``switching" node $v^*(\bar{y}^n)$. We argue that if $|s_i(\bar{y}^{i}) |> n \mu$ for some $i \in\{1,\ldots, k-1\}$ or $|s_k(\bar{y}^{k})| \leq n \mu$, then  we can find a new ordered tree $\mathcal{T}'$  with a larger set of leaves exceeding the desired score than the original tree. I.e.,  $\mathcal{S}(\mathcal{T}')\supseteq \mathcal{S}(\mathcal{T})$ which by \eqref{eq:PT}  implies $P(\mathcal{T}')\geq P(\mathcal{T})$. It can also be verified that the procedure will produce $\mathcal{T}^*$ after a finite number of steps. The new ordered tree $\mathcal{T}'$ is obtained as follows:
    \begin{itemize}
        \item Assume $|s_i(\bar{y}^{i})| > n \mu$ for some $i \in\{1,\ldots, k-1\}$. In this case, denote the level-$i$ node on the path to $\bar{y}^n$ by $\tilde{v}$ and set in $\mathcal{T}'$ the new node $\tilde{v}$ as  the ``switching" node of $y^n$ 
        (and naturally of all descendant leaves of $\tilde{v}$). Notice that all descendant leaves $y^n$ of $\tilde{v}$ have score $|s_n(y^n, \mathcal{T}')|> n \mu$  implying in particular that this holds for all  leaves that are descendants of the original ``switching" node 
        $v^*(\bar{y}^n)$. Since the score of all other leaves did not change from $\mathcal{T}$ to $\mathcal{T'}$, we can conclude the  inclusion $\mathcal{S}(\mathcal{T}')\supseteq \mathcal{S}(\mathcal{T})$.
        \item If $|s_k(\bar{y}^{k}) | \leq n \mu$, then all leaves $y^n$ that are descendants of $v^*(\bar{y}^n)$  has scores below $n \mu$ and do not belong to $\mathcal{S}(\mathcal{T})$. We construct  $\mathcal{T}'$ from $\mathcal{T}$ by relabeling 
        $v^*(\bar{y}^n)$  by $a$, and trivially conclude  $\mathcal{S}(\mathcal{T}') \supseteq\mathcal{S}(\mathcal{T})$. 
    \end{itemize}
    \end{IEEEproof}
\subsubsection{Success Probability  of   Strategy $\mathbf{h}^*$ in \eqref{eq:optimal} }

\begin{prop}We have:
    \begin{equation}
 \PP\left( |S_n(\mathbf{h}^*,a,b) |>n \mu\right)\leq\dfrac{1}{4 n\mu^2}
    \end{equation}
\end{prop}

\begin{IEEEproof}Define the i.i.d. variables $ Z_k =  \indic\{\tilde{Y}_i=b\} - P_{Y|X}(b|a)$ for $k=1,2,\ldots$ where   $\{\tilde{Y}_i\}_{i=1}^{+\infty}$  is i.i.d.  $\sim P_{Y|X}(\cdot|a)$. 
 By Kolmogorov's maximal inequality~\cite[Theorem 2.5.5]{Durretbook}:
    \begin{IEEEeqnarray}{rCl}
        \PP\left(\max_{1\leq k\leq n}\left|\sum_{i=1}^k Z_i\right| > n\mu\right)&\leq& \dfrac{\EE[(\sum_{k=1}^n Z_k)^2]}{n^2\mu^2}
         \leq \frac{1}{4n \mu^2}.\label{eq:bound}\IEEEeqnarraynumspace
    \end{IEEEeqnarray}
    
We next relate the  probability on the left-hand side of \eqref{eq:bound} with the success probability of our optimal strategy $\mathbf{h}^*$. Notice that for all $y^n$ sequences where  $\max_{1\leq k\leq n} |\sum_{i\leq k}Z_i| \leq  n\mu$ when $\tilde{Y}^n=y^n$, we also have $|s_k(y^k, \mathbf{h}^*)| \leq n \mu$ for all $k\in\{1,\ldots, n\}$. This is,  \emph{both are \emph{smaller} than the threshold.} 
Moreover, these sequences have same probability of arising for the random sequences $\tilde{Y}^n$  and $Y^n$ as the optimal strategy $\mathbf{h}^*$ always produces $a$, i.e., $x_1=\ldots=x_n=a$. 
We  obtain: 
\begin{equation} 
        \PP\left(\max_{1\leq k\leq n}\left |\sum_{i=1}^k Z_i\right| > n\mu\right)\leq \PP\left( |S_{n}(Y^n, \mathbf{h}^*)| \leq n\mu\right). 
\end{equation} 
The proof is then concluded by considering the complement probabilities as well as \eqref{eq:bound}.

   \end{IEEEproof}

\section{Converse Theorem for an ISAC Model}
\begin{figure}
    \centering
    \includegraphics[scale=0.7]{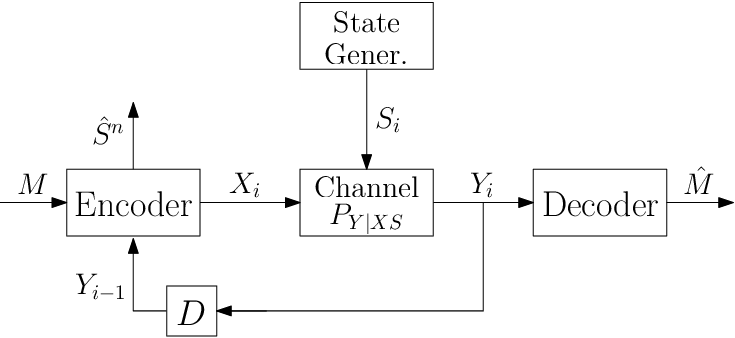}
    \caption{ISAC Model}
    \label{ISAC}
\end{figure}

We use our Lemma~\ref{lemmimp} to prove a converse for the  integrated sensing and communication (ISAC) model in Figure~\ref{ISAC}.  

A transmitter seeks to communicate a random message $M$ to a receiver over a state-dependent discrete memoryless channel (SDMC) $P_{Y|XS}$. The message $M$ is assumed to be uniformly distributed over the set $\llbracket 1,2^{nR} \rrbracket$, where $R > 0$ denotes the communication rate and $n > 0$ the blocklength.
The SDMC is influenced by a state sequence $S^n$, which is i.i.d. according to a given probability mass function (pmf) $P_S$. At each time $i$, the channel output $Y_i$ is generated based on the current input $X_i$ and the state $S_i$, according to the SDMC transition probability $P_{Y|XS}$. The outputs are observed at the receiver, and through backscattering also at the transmitter. 

The transmitter creates its channel inputs based on the message and the past feedback outputs
\begin{equation}
    X_i = f_i^{(n)}(M, Y_1, \ldots, Y_{i-1}), \qquad i = 1, \ldots,  n,
\end{equation}
for some sequence of encoding functions $\{f_i^{(n)}\}_{=1}^n$. It also uses the feedback outputs $Y^n$ and the produced input sequence $X^n$ (or $M$) to reconstruct the channel's state sequence $S$:
\begin{equation}
    \hat{S}^n = h^{(n)}(X^n, Y^n).
\end{equation}

The receiver attempts to guess the original message $M$ based on the full sequence of channel outputs $Y^n$ using a decoding function $g^{(n)}$: 
\begin{equation}
    \hat{M} = g^{(n)}(Y^n).
\end{equation}

Reliability of  communication  is evaluated via the average probability of decoding error at the receiver
\begin{equation}
    P_e^{(n)} = \PP\left(\hat{M} \neq M\right).
\end{equation}
Sensing performance at the transmitter is 
measured by the average expected distortion between the true state sequence $S^n$ and its reconstruction $\hat{S}^n$:\begin{equation}
    \text{dist}^n(\hat{S}^n, S^n) = \frac{1}{n} \sum_{i=1}^n d(\hat{S}_i, S_i),
\end{equation}
where $d$ denotes a given positive and bounded distortion function.

Our goal is to have small decoding error probability and small excess distortion probability. 
\begin{defi}
    A rate-distortion pair $(R,D)$ is $(\eps, \delta)$-achievable if there exist sequences of encoding, decoding and estimation functions $\{\{f_k^{(n)}\}_{k=1}^n,g^{(n)},h^{(n)}\}_{n=1}^{+\infty}$ satisfying
    \begin{IEEEeqnarray}{rCl}
        \label{ISACerror}
        \varlimsup_{n\rightarrow+\infty}P_e^{(n)} &\leq& \eps \\
        \label{ISACexcess}
        \varlimsup_{n\rightarrow\infty} \PP\left(\text{dist}^n\left(\hat{S}^n,S^n\right) > D\right)&\leq &\delta.
    \end{IEEEeqnarray}
\end{defi}

\begin{theo}
    \label{theoISAC}
    For any $\eps +\delta <1$, if a rate-distorsion pair $(R,D)$ is $(\eps,\delta)$-achievable then there exists $P_X$ satisfying
    \begin{IEEEeqnarray}{rCl}        R&=&I_{P_X P_S P_{Y|XS}}(X;Y)\\
        D&\geq&\EE_{P_X P_S P_{Y|XS}}\left[d(\hat{s}(X,Y),S)\right]
    \end{IEEEeqnarray}
    where
    \begin{equation}\label{eq:shat}
        \hat{s}(x,y)=\min_{\hat{s}\in\mathcal{S}}\sum_s P_{S|XY}(s|x,y)d(\hat{s},s).
    \end{equation}
\end{theo}
Achievability for $\epsilon= \delta=0$ follows directly from the results in  \cite{kobayashi2018joint,b6}, where it was also shown that the transmitter does not need to rely on the feedback outputs $Y_{1}, \ldots, Y_{i-1}$ to produce the next input $X_i$.  The contribution of our article lies in  proving the converse to this statement in the case where the encoder can use the feedback to produce the channel inputs. 
\begin{remark}\label{rem:range}
Notice that under the average probability of error criterion considered in this article, above theorem does not hold when $\epsilon+\delta \geq 1$. Better performances can be achieved. 
\end{remark} 

\begin{remark} 
Theorem~\ref{theoISAC} and Remark~\ref{rem:range} can easily be extended also to setups where the transmitter observes a (deterministically or stochastically) degraded version of the outputs $\{Y_i\}$. 
\end{remark}
\subsection{Proof of Theorem \ref{theoISAC}}

Fix $\epsilon, \delta>0$ with sum $\epsilon+\delta<1$. Fix also a sequence of encoding  and decoding functions $\{\{f_i^{(n)}\}_{i=1}^n,g^{(n)}\}_{n=1}^\infty$ satisfying   (\ref{ISACerror}) and (\ref{ISACexcess}) and consider the optimal state estimator  (see \cite[Lemma 1]{b6} $h^n(s^n,x^n)=(\hat{s}(x_1,y_1), \ldots, \hat{s}(x_n,y_n))$, where $\hat{s}(x,y)$ is defined in \eqref{eq:shat}. 

Define for any blocklength $n$ the parameter $ \mu_n=n^{-1/4}$ and notice that $\mu_n \to 0$ while $n \mu_n^2\to \infty$ as $n\to +\infty$.
Consider now the three conditions:
\begin{align}
    g^{(n)}(y^n) &= m \label{ISACcond1} \\
    \text{dist}^n(h^{(n)}(x^n,y^n),s^n) &\leq D \label{ISACcond2} \\
    \left| \pi_{x^n,s^n,y^n}(a,b,c) - \pi_{x^n}(a)P_S(b)P_{Y|XS}(c|a,b) \right| &\leq \mu_n \label{ISACcond3}
\end{align}
where in the above, $x_i=f_i(m,y^{i-1})$ for any $i=1,\ldots, n$. 

Fix a constant $\eta$ such that $0 < \eta < 1 - \varepsilon - \delta$, and define the subset $\tilde{\mathcal{M}}_n \subseteq \llbracket 1, 2^{nR} \rrbracket$ to consist of all messages $m$ satisfying 
\begin{IEEEeqnarray}{rCl}
    \PP\left( \tilde{M} \neq M  \textnormal{ or } \mathrm{dist}^n\left( \hat{S}^n, S^n \right) > D \mid M = m \right) & \leq & 1 - \eta, \IEEEeqnarraynumspace\label{eq:subexp1}
\end{IEEEeqnarray}
and let $\tilde{M}$  be uniformly distributed  over $\tilde{\mathcal{M}}_n$. 
Notice that  \eqref{ISACerror} and \eqref{ISACexcess} imply that (see Appendix~\ref{app:D}): 
\begin{equation}
    \label{eq:cardMtilden}
    \frac{|\tilde{\mathcal{M}}_n|}{2^{nR}} \geq \left(1 - \frac{P_e^{(n)} + P_{D}^{(n)}}{1 - \eta}\right) =: \gamma_n,
\end{equation}
where we define  $P_D^{(n)}:=\PP\left(\text{dist}^n\left(\hat{S}^n,S^n\right) > D\right)$. 

Define now for any $m\in\tilde{\mathcal{M}}_n$ the set $\mathcal{D}_{n,m}$ consisting of all pairs $(s^n, y^n)$ such that (\ref{ISACcond1}), (\ref{ISACcond2}), and (\ref{ISACcond3}) hold, 
and introduce random variables $( \tilde{S}^n, \tilde{Y}^n)$ jointly distributed with $\tilde{M}$ through the conditional pmfs:
\begin{IEEEeqnarray}{rCl}
 \lefteqn{   P_{\tilde{S}^n,\tilde{Y}^n| \tilde{M}=m}(s^n,y^n) } \hspace{1cm}\nonumber \\
 &\triangleq & 
    \frac{P_S^{\otimes n}(s^n) \prod_{i=1}^n  P_{Y|XS}(y_i |x_i(m,y^{i-1}),s_i)}{\Delta_{n,m}} \nonumber\\
    && \cdot  \indic\{ (s^n, y^n) \in \mathcal{D}_{n,m} \},
\end{IEEEeqnarray}
where $\Delta_{n,m} \triangleq \PP \left( (S^n, Y^n) \in \mathcal{D}_{n,m} \,|\, M=m\right)$. 

Notice that:
    \begin{IEEEeqnarray}{rCl}
    \lefteqn{\Delta_{n,m}}\nonumber \\
     &= &\; 1-\PP\left((S^n,Y^n)\notin \mathcal{D}_{n,m}\,| M=m\right) \\ 
      &           = & \;1-\PP(M\neq \hat{M} \text{  or  } \text{dist}^n(\hat{S}^n, S^n)>D \text{  or  }\nonumber \\
        &     &\hspace{9mm} \exists(a,b,c)\colon | \pi_{X^n,Y^n,S^n}(a,b,c)\nonumber \\
          &   & \hspace{6mm} -\pi_{X^n}(a)P_S(c)P_{Y|X}(b|a)|>\mu_n\,|\, M=m)\IEEEeqnarraynumspace\\
     &       \geq & \; \eta  - \sum_{a,b,c} \PP\left(|\pi_{X^n,Y^n,S^n}(a,b,c)\right.\nonumber \\
                 & & \qquad \left.-\pi_{X^n}(a)P_{Y|X}(b|a)P_S(c)|>\mu_n\, |\, M=m\right)\IEEEeqnarraynumspace
\label{eq:a}
\end{IEEEeqnarray}
By  Lemma \ref{lemmimp}, we obtain  for any $m$ and triple $(a,b,c)$:
\begin{IEEEeqnarray}{rCl}
\lefteqn{\PP\Big(|\pi_{X^nY^nS^n}(a,b,c) } \nonumber\\
&& \hspace{.6cm} -\pi_{X^n}(a)P_{Y|X}(b|a)P_S(c)|>\mu_n \big| {M}=m\Big)  \leq \frac{1}{4 n\mu_n^2}, \IEEEeqnarraynumspace\label{eq:b}
\end{IEEEeqnarray}
and thus, by \eqref{eq:a}:
\begin{equation}
    \label{eq:upper_bound_Delta}
    \Delta_{n,m} \geq \eta-\dfrac{|\mathcal{X}| |\mathcal{Y}| |\mathcal{S}|}{4n\mu_n^2}, \quad \forall m\in\tilde{\mathcal{M}}_n.
\end{equation}

We now prove the desired bounds on the rate and the distortion. By combining~(\ref{eq:cardMtilden}) and~(\ref{ISACcond1}), we establish the following inequality, the proof of which is provided in Appendix~\ref{app:ineqR}:

\begin{IEEEeqnarray}{rCl}
    \label{eq:ISACravantlimite}
    R \leq H(\tilde{Y}_{T_n}) - \dfrac{1}{n} H(\tilde{Y}^n | \tilde{M}) + o(1), 
\end{IEEEeqnarray}
where we defined  $T_n$ to be independent of $(\tilde{M},\tilde{X}^n,\tilde{Y}^n)$ and uniform over $\llbracket 1,n\rrbracket$.

\begin{lemm}
    \label{lemm4}
    There exists a subsequence $\{n_i\}_{i \geq 1}$ and a pmf $P_X$ on $\mathcal{X}$ such that:
\begin{IEEEeqnarray}{rCl}
         \lim_{i\rightarrow +\infty} P_{\tilde{X}_{T},\tilde{Y}_{T}, \tilde{S}_{T}}(x,y,s)&=&P_X(x)P_S(s)P_{Y|XS}(y|x,s) \label{eq:tildeISACcond2} \IEEEeqnarraynumspace\\
        \varliminf_{i \rightarrow +\infty} \dfrac{1}{n_i} H(\tilde{Y}^{n_i} | \tilde{M}) &\geq& H(Y | X)\label{H}
\end{IEEEeqnarray}
    where the random variables $(X, Y)$ in the entropy term are distributed according to the  joint marginal of pmf $P_X P_S P_{Y|XS}$.
\end{lemm}
\begin{IEEEproof} 
Based on~(\ref{ISACcond3}) and~(\ref{eq:upper_bound_Delta}), see Appendix~\ref{proof:lemma3}.
\end{IEEEproof}

Subsequently, by virtue of this lemma and~(\ref{eq:ISACravantlimite}), and as the blocklength $n_i$ tends to $+\infty$, it must hold that:
\begin{equation}
    R\leq H(Y)-H(Y|X) = I(X;Y).
\end{equation}

We prove  the distortion bound. By  (\ref{ISACcond2}), with probability 1:
\begin{IEEEeqnarray}{rCl}    D&\geq &\dfrac{1}{n_i}\sum_{j=1}^{n_i} d\left(\hat{s}\left( \tilde{X}_j, \tilde{Y}_j\right),\tilde{S}_j\right)\\
      & = & \EE\left[d\left(\hat{s}\left( \tilde{X}_{T_n}, \tilde{Y}_{T_n}\right),\tilde{S}_{T_n}\right)\right]
\end{IEEEeqnarray}
and by (\ref{eq:tildeISACcond2}), we can conclude by taking  $n_i\to +\infty$ that:
\begin{equation}
    D\geq \EE\left[d(\hat{s}(X,Y),S)\right].
\end{equation}

\section{Conclusion}

In this paper, we analyzed a closed-loop communication system where the encoder adapts its transmissions based on previous outputs of a discrete memoryless channel. We established that, despite the presence of feedback, the conditional type of the channel outputs given the inputs remains close to the true channel law \( P_{Y|X} \). The core of the proof involved identifying the worst-case encoder strategy and bounding the probability of significant deviation using a random walk argument.

We then applied this general result to an Integrated Sensing and Communication (ISAC) model, where a transmitter seeks to jointly communicate a message and estimate the state of a channel based on the backscattered feedback signals. By leveraging the conditional type result and a change-of-measure technique, we proved that the fundamental limits are the same as in the corresponding open loop systems with non-adaptive channel inputs.


\clearpage
\appendices

\section{Proof of the Inequality in \eqref{eq:ineq}}\label{app:same}


Consider  the tree $\mathcal{T}_{k-1}$, the chosen node $v_k$, and the constructed random tree  $\tilde{\mathcal{A}}$. Then, let $\ell$ be the depth of node $v_k$ and
for each $y\in \mathcal{Y}$, let $\mathcal{A}_y$ be the realization of the tree $\tilde{\mathcal{A}}$ that is constructed using the augmented version of $\mathcal{B}_y$ and which occurs with probability $P_{Y|X}(y|\tilde{a})$.

Notice first that  any descendant leaf $y^n$ of node $v_k$,  has depth- $\ell+1$ score $\mathbb{1}\{x_{\ell+1}=a\}\left( \mathbb{1}\{y_{\ell+1}=b\}-P_{Y|X}(b|a)\right)$ equal 0 in tree  $\mathcal{T}_{k-1}$ because $v_k$ is labeled $\tilde{a}\neq a$. Similarly, any descendant leaf $\tilde{y}^n$ of node $v_k$, has depth-$n$ score $\mathbb{1}\{x_{n}=a\}\left( \mathbb{1}\{y_{n}=b\}-P_{Y|X}(b|a)\right)$  equal to 0 in tree $\mathcal{A}_{y}$ because the depth-$n-1$ node of this leaf is labeled $\tilde{a}$. We thus have for any descendant leaf $\tilde{y}^n$ of  node $v_k$:
\begin{IEEEeqnarray}{rCl} 
s_n(\tilde{y}^n, \mathcal{A}_{y'})& =  & s_{n-1}(\tilde{y}^{n-1}, \mathcal{A}_{y'}) \\
&= &  s_n((\tilde{y}^{\ell}, y', \tilde{y}_{\ell+1}^{n-1}), \mathcal{T}_{k-1}),\label{eq:d}
\end{IEEEeqnarray}
where in the last equality we used that leaf ${y}^n= (\tilde{y}^{\ell-1}, y', \tilde{y}_{\ell}^{n-1})$ is also a descendant  of depth-$\ell$ node  $v_k$. 

We can use this equality to calculate the \emph{expected} success probability of the tree $\tilde{\mathcal{A}}$ where expectation is over the random tree construction. For any tree, let $x_i(y^{i-1},\mathcal{T})$ denote the label in tree  $\mathcal{T}$ of the node that is reached from the root over the path $y^{i-1}$. We can then write: 
\begin{IEEEeqnarray}{rCl} 
\lefteqn{\EE[ P(\tilde{\mathcal{A}})]}  \nonumber\\
 & = & \sum_{y' \in \mathcal{Y}}P(A_{y'})P_{Y|X}(y'|\tilde{a}) \\
&= &  \sum_{y' \in \mathcal{Y}} \sum_{\tilde{y}^n \in \mathcal{S}(\mathcal{A}_{y'})}  \prod_{i=1}^n P_{Y|X}\big(\tilde y_i| x_i(\tilde y^{i-1},\mathcal{A}_{y'})\big)P_{Y|X}(y'|\tilde{a}). \nonumber\\\label{eq:overall}
\end{IEEEeqnarray}
We now partition the set $\mathcal{S}(\mathcal{A}_{y'})$ into leaves that are descendants of $v_k$, denoted $\mathcal{S}_{v_k}(\mathcal{A}_{y'})$ and those that are not, denoted $\mathcal{S}^{c}_{v_k}(\mathcal{A}_{y'})$. We then observe that irrespective $y'$:
\begin{IEEEeqnarray}{rCl} 
\lefteqn{\sum_{\tilde{y}^n \in \mathcal{S}_{v_k}^{c}(\mathcal{A}_{y'})}  \prod_{i=1}^n P_{Y|X}\big(\tilde y_i| x_i(\tilde y^{i-1},\mathcal{A}_{y'})\big)} \nonumber\qquad  \\
&= & \sum_{\tilde{y}^n \in \mathcal{S}_{v_k}^{c}(\mathcal{T}_{k-1})  }\prod_{i=1}^n P_{Y|X}\big(\tilde y_i| x_i(\tilde y^{i-1},\mathcal{T}_{k-1})\big),\label{eq:1}
\end{IEEEeqnarray}
because $ \mathcal{S}_{v_k}^{c}(\mathcal{A}_{y'}) = \mathcal{S}_{v_k}^{c}(\mathcal{T}_{k-1})$ as nothing has changed on the labels on the paths from the root to these leaves. 

On the other hand, 
\begin{IEEEeqnarray}{rCl} 
\lefteqn{\sum_{\tilde{y}^n \in \mathcal{S}_{v_k}(\mathcal{A}_{y'})}  \prod_{i=1}^n P_{Y|X}\big(\tilde y_i| x_i(\tilde y^{i-1},\mathcal{A}_{y'})\big)} \nonumber\\
&\stackrel{(a)}{=}& \sum_{\tilde{y}^n \in \mathcal{S}_{v_k}(\mathcal{A}_{y'})}  \prod_{i=1}^{n-1} P_{Y|X}\big(\tilde y_i| x_i(\tilde y^{i-1},\mathcal{A}_{y'})\big)P_{Y|X}(\tilde{y}_n | \tilde{a})\\
&\stackrel{(b)}{=} & \sum_{ \substack{\tilde{y}^{n-1} \colon\\\ (\tilde y^{n-1},y') \in \mathcal{S}_{v_k}(\mathcal{A}_{y'})}}  \prod_{i=1}^{n-1} P_{Y|X}\big(\tilde y_i| x_i(\tilde y^{i-1},\mathcal{A}_{y'})\big) \\
&\stackrel{(c)}{=}& \sum_{ \substack{\tilde{y}^{n-1} \colon\\\ (\tilde y^{n-1},y') \in \mathcal{S}_{v_k}(\mathcal{A}_{y'})}}   \prod_{i=1}^{\ell} P_{Y|X}\big(\tilde y_i| x_i(\tilde y^{i-1},\mathcal{T}_{k-1})\big) \nonumber \\
&& \hspace{1.8cm} \cdot \prod_{i=\ell+1}^{n-1} P_{Y|X}\big(\tilde y_i| x_i((\tilde y^{\ell}, y',\tilde{y}_{\ell+1}^{i-1}),\mathcal{T}_{k-1})\big)\\
&\stackrel{(d)}{=}& \sum_{\substack{{y}^n \in \mathcal{S}_{v_k}(\mathcal{T}_{k-1}) \colon\\
y_\ell =y' }} \prod_{i=1}^n P_{Y|X}\big( y_i| x_i( y^{i-1},\mathcal{T}_{k-1})\big) \frac{1}{P_{Y|X} (y{'}|\tilde{a})},\nonumber\\
 \label{eq:2}
\end{IEEEeqnarray}
where  $(a)$ holds because in tree $\mathcal{A}_{y'}$ all depth-$(n-1)$ descendants of $v_k$ are labeled $\tilde{a}$; $(b)$ holds because if $\tilde{y}^n$ belongs to $\mathcal{S}_{v_k}(\mathcal{A}_{y'})$, so do all siblings of $\tilde{y}$, and because $\sum_{\tilde{y}_n} P_{Y|X}(\tilde{y}_n|\tilde{a})=1$; $(c)$ holds by the construction of tree $\mathcal{A}_{y'}$ which inherits all labels from the root to the path of $v_k$ and then jumps to the subtree along edge $y'$; and $(d)$ holds by \eqref{eq:d} and because $v_k$ is labeled $\tilde{a}$.  

Plugging \eqref{eq:1} and \eqref{eq:2} into \eqref{eq:overall}, we finally obtain: 
\begin{IEEEeqnarray}{rCl} 
\lefteqn{\EE[ P(\tilde{\mathcal{A}})]}  \nonumber\\
&= &  \sum_{y' \in \mathcal{Y}}P_{Y|X}(y'|\tilde{a})  \sum_{\tilde{y}^n \in \mathcal{S}_{v_k}^{c}(\mathcal{T}_{k-1})  }\prod_{i=1}^n P_{Y|X}\big(\tilde y_i| x_i(\tilde y^{i-1},\mathcal{T}_{k-1})\big)  \nonumber \\
&& +   \sum_{y' \in \mathcal{Y}} \; \sum_{\substack{{y}^n \in \mathcal{S}_{v_k}(\mathcal{T}_{k-1}) \colon\\
y_\ell =y' }} \prod_{i=1}^n P_{Y|X}\big( y_i| x_i( y^{i-1},\mathcal{T}_{k-1})\big) \\
 & =&  \sum_{\tilde{y}^n \in \mathcal{S}_{v_k}^{c}(\mathcal{T}_{k-1})  }\prod_{i=1}^n P_{Y|X}\big(\tilde y_i| x_i(\tilde y^{i-1},\mathcal{T}_{k-1})\big) \nonumber\\
 && + \sum_{{y}^n \in \mathcal{S}_{v_k}(\mathcal{T}_{k-1})} \prod_{i=1}^n P_{Y|X}\big( y_i| x_i( y^{i-1},\mathcal{T}_{k-1})\big)\\
 &= & P( \mathcal{T}_{k-1}). 
\end{IEEEeqnarray}
This concludes the proof. 
\section{Proof of Bound \eqref{eq:cardMtilden}}\label{app:D}
Recall the definitions $P_e^{(n)}=\PP(M\neq \hat{M})$ and  $P_D^{(n)}=\PP\left(\text{dist}^n\left(\hat{S}^n,S^n\right) > D\right)$. 

By the union bound, we have
\begin{IEEEeqnarray}{rCl} 
\lefteqn{\EE_{M}\left[  \PP\left[ \tilde{M} \neq M \; \textnormal{or}\; \text{dist}^n\left(\hat{S}^n,S^n\right) > D  \mid M\right] \right]} \qquad \nonumber\\
&  \leq & P_e^{(n)}+P_D^{(n)}. \hspace{3cm}
\end{IEEEeqnarray}
Considering the conditional probability in above equation as a nonnegative function of $M$ (and thus a new random variable), by Markov's 
 inequality we have: 
\begin{IEEEeqnarray}{l} 
\PP_M\left[   \PP\left[ \tilde{M} \neq M \; \textnormal{or}\; \text{dist}^n\left(\hat{S}^n,S^n\right) > D] >1-\eta \mid M \right] \right] \nonumber\\
  \leq  \frac{ P_e^{(n)}+P_D^{(n)}}{1-\eta}. 
\end{IEEEeqnarray}
The desired inequality then follows by noting that the left-hand side of above inequality is equal to $1- \frac{|\tilde{\mathcal{M}}_n|}{2^{nR}}$ since all messages have probability $2^{-nR}$.

\section{Proof of the Inequality~(\ref{eq:ISACravantlimite})}\label{app:ineqR}

By definition of $\tilde{M}$ and~(\ref{eq:cardMtilden})
\begin{equation}
    P_{\tilde{M}}(m)=\dfrac{1}{|\tilde{\mathcal{M}}_n|}\leq\dfrac{1}{\gamma_n 2^{nR}}, \quad \forall m\in\tilde{\mathcal{M}}_n,
\end{equation}
thus
\begin{IEEEeqnarray}{rCl}
    H(\tilde{M}) & = & -\sum_{m\in\tilde{\mathcal{M}}_n}P_{\tilde{M}}(m)\log\left(P_{\tilde{M}}(m)\right) \\
    & \geq & \sum_{m\in\tilde{\mathcal{M}}_n}P_{\tilde{M}}(m)\log\left(\gamma_n 2^{nR}\right) \\
    & = & \log(\gamma_n) +nR
\end{IEEEeqnarray}

We then obtain the upper bound for $R$ :
\begin{IEEEeqnarray}{rCl}
    R & \leq & \dfrac{1}{n}H(\tilde{M}) -\dfrac{1}{n}\log(\gamma_n)\\
     & \stackrel{(a)}{=} & \dfrac{1}{n}I(\tilde{Y}^n;\tilde{M})+o(1) \\
     & = & \dfrac{1}{n}H(\tilde{Y}^n) - \dfrac{1}{n}H(\tilde{Y}^n \,|\, \tilde{M}) +o(1)
\end{IEEEeqnarray}
where $(a)$ holds because $\tilde{M}$ is a deterministic function of $\tilde{Y}^n$ as~(\ref{ISACcond1}) imposes that there is no decoding error $\tilde{M}=g^{(n)}(\tilde{Y}^n)$ and because $\dfrac{1}{n}\log(\gamma_n)$ vanishes for increasing blocklengths by its definition,  by assumptions  (\ref{ISACerror}) and (\ref{ISACexcess}), and because $\epsilon+\delta < 1-\eta$. 

Continue to consider the first term:
\begin{IEEEeqnarray}{rCl}
    \dfrac{1}{n}H(\tilde{Y}^n) & = & \dfrac{1}{n}\sum_{i=1}^n H(\tilde{Y}_i| \tilde{Y}^{i-1}) \\
    & \leq & \dfrac{1}{n}\sum_{i=1}^nH(\tilde{Y}_i) \\
    & = & \sum_{i=1}^n P_{T_n}(i)H(\tilde{Y}_T\, |\, T=i)\\
    & = & H(\tilde{Y}_T\,|\, T) \\
    & \leq & H(\tilde{Y}_T)
\end{IEEEeqnarray}
This concludes the proof of the inequality.

\section{Proof of  Lemma \ref{lemm4}}\label{proof:lemma3}

To prove the first limit \eqref{eq:tildeISACcond2} we fix a triple $(x,y,s)\in\mathcal{X} \times \mathcal{Y}\times \mathcal{S}$ and $n \geq 1$. Then, 
    \begin{IEEEeqnarray}{rCl}
    \lefteqn{    P_{\tilde{X}_{T_n}\tilde{Y}_{T_n}\tilde{S}_{T_n}}(x,y,s) } \nonumber \qquad \\
    & = &\dfrac{1}{n}\sum_{i=1}^n P_{\tilde{X}_i\tilde{Y}_i\tilde{S}_i}(x,y,s)\\
        & =&  \EE\left[\dfrac{1}{n}\sum_{i=1}^n \indic\{\tilde{X}_i=x. \tilde{Y}_i=y,\tilde{S}_i=s\} \right] \\
        & = &\EE[\pi_{\tilde{X}^n \tilde{Y}^n,\tilde{S}^n}(x,y)] \\
        & \leq   &      \pi_{\tilde{X}^n}(x)P_{Y|XS}(y|x,s)P_S(s) + \mu_n,\label{eq:upper}
    \end{IEEEeqnarray}
    where the inequality holds by Condition~\eqref{ISACcond2}. In a similar way, one can also show that 
    \begin{align}
        P_{\tilde{X}_{T_n}\tilde{Y}_{T_n}\tilde{S}_{T_n}}(x,y,s) 
        & \geq         \pi_{\tilde{X}^n}(x)P_{Y|XS}(y|x,s)P_S(s) - \mu_n.\label{eq:lower}
    \end{align}
    Notice that $\{\EE[\pi_{\tilde{X}^n}(x)]\}_{n \geq 1}$ is a bounded sequence, so by the Bolzano-Weierstrass theorem, there exists a subsequence $\{n_i\}$ and an accumulation point $P_X$ such that:
    \begin{equation}
        \lim_{i \rightarrow +\infty} \EE[\pi_{\tilde{X}^{n_i}}(x)] = P_X(x), \qquad \forall x \in \mathcal{X}.
    \end{equation}
 Combining  \eqref{eq:upper}--\eqref{eq:lower}, since  $\mu_n\to0$ as $n\to +\infty$, proves the first limit \eqref{eq:tildeISACcond2}. 

    To prove the second limit \eqref{H}, fix $m \in \tilde{\mathcal{M}}_n$ and define:
    \begin{equation}
        \mathcal{D}_{n,m}' = \{ y^n \in \mathcal{Y}^n \,|\, \exist s^n \colon (s^n, y^n) \in \mathcal{D}_{n,m} \},
    \end{equation}
    Then:
    \begin{IEEEeqnarray}{rCl}
        \lefteqn{\dfrac{1}{n} H(\tilde{Y}^n |\tilde{M} = m)} \nonumber \\
        & = & -\dfrac{1}{n}\sum_{y^n \in \mathcal{D}_{n,m}'} P_{\tilde{Y}^n |\tilde{M} = m}(y^n)\log{(P_{\tilde{Y}^n |\tilde{M} = m}(y^n))} \\
        & \geq & -\dfrac{1}{n}\sum_{y^n \in \mathcal{D}_{n,m}'} P_{\tilde{Y}^n |\tilde{M} = m}(y^n)\log{\left(\dfrac{P_{Y^n | M = m}(y^n)}{\Delta_{n,m}}\right)} \\
        & = & -\dfrac{1}{n}\sum_{y^n \in \mathcal{D}_{n,m}'} P_{\tilde{Y}^n |\tilde{M} = m}(y^n) \nonumber \\
        && \hspace{0.3cm} \cdot \sum_{i=1}^n \log{(P_{Y_i | M, Y^{i-1}}(y_i | m, y^{i-1}))} + \dfrac{\log(\Delta_{n,m})}{n} \\
        & \stackrel{(a)}{\geq} & -\dfrac{1}{n}\sum_{y^n \in \mathcal{D}_{n,m}'} P_{\tilde{Y}^n |\tilde{M} = m}(y^n) \nonumber \\
        & & \hspace{1cm} \cdot \sum_{i=1}^n \log{(P_{Y|X}(y_i | x_i(m,y^{i-1})))} \nonumber \\
        & & \hspace{0cm} + \dfrac{1}{n}\log\left(\eta-\dfrac{|\mathcal{X}| |\mathcal{Y}| |\mathcal{S}|}{4n\mu_n^2}\right) \\
        & = & -\dfrac{1}{n}\sum_{y^n \in \mathcal{D}_{n,m}'} \sum_{i=1}^n P_{\tilde{Y}^n |\tilde{M} = m}(y^n) \nonumber \\
        & &  \hspace{1cm} \cdot \sum_{a,b} \indic\{x_i(m,y^{i-1}) = a, y_i = b\} \log{(P_{Y|X}(b|a))} \nonumber \\
        && \hspace{0cm} +\dfrac{1}{n}\log\left(\eta-\dfrac{|\mathcal{X}| |\mathcal{Y}| |\mathcal{S}|}{4n\mu_n^2}\right) \\
        & = & -\sum_{y^n \in \mathcal{D}_{n,m}'} P_{\tilde{Y}^n |\tilde{M} = m}(y^n)  \nonumber \\
        && \hspace{1cm} \cdot \left(\sum_{a,b} \pi_{x^n(m,y^n), y^n}(a,b) \log{(P_{Y|X}(b|a))} \right)\nonumber \\
        & & \hspace{0cm} +\dfrac{1}{n}\log\left(\eta-\dfrac{|\mathcal{X}| |\mathcal{Y}| |\mathcal{S}|}{4n\mu_n^2}\right),\label{eq:last}
    \end{IEEEeqnarray}
    where $(a)$ holds by~(\ref{eq:upper_bound_Delta}). 
        
  Notice next that by summing Condition~(\ref{ISACcond3}) over $\mathcal{S}$, we obtain inequality
    \begin{IEEEeqnarray}{rCl}
  \pi_{x^n(m,y^n), y^n}(a,b)& \geq &         \pi_{x^n(m,y^n)}(a) P_{Y|X}(b|a) - |\mathcal{S}|\mu_n.\label{eq:2p}\IEEEeqnarraynumspace
     \end{IEEEeqnarray}

By~(\ref{eq:last}) and~(\ref{eq:2p}), we may deduce that:
    \begin{IEEEeqnarray}{rCl}
        \lefteqn{\dfrac{1}{n} H(\tilde{Y}^n |\tilde{M} = m)} \qquad \nonumber \\
        &\geq & \sum_{y^n \in \mathcal{D}_{n,m}'} P_{\tilde{Y}^n |\tilde{M} = m}(y^n) \nonumber \\
        && \hspace{1cm} \cdot \left(\sum_{a \in \mathcal{X}} \pi_{x^n(m,y^n)}(a) H(Y|X = a)\right) 
         \nonumber\\
        && + \sum_{\substack{(a, b)\colon \\P_{Y|X}(b|a)>0}} |\mathcal{S}|\mu_n  \log P_{Y|X}(b|a)
        \nonumber \\
        & & \hspace{0cm} +\dfrac{1}{n}\log\left(\eta-\dfrac{|\mathcal{X}| |\mathcal{Y}| |\mathcal{S}|}{4n\mu_n^2}\right)\\
        & = & \sum_{a \in \mathcal{X}} H(Y|X = a) \EE[\pi_{\tilde{X}^n}(a) | \tilde{M} = m] \nonumber\\
        && + \sum_{\substack{(a, b)\colon \\P_{Y|X}(b|a)>0}} |\mathcal{S}| \mu_n \log P_{Y|X}(b|a)
        \nonumber \\
        & & \hspace{0cm} +\dfrac{1}{n}\log\left(\eta-\dfrac{|\mathcal{X}| |\mathcal{Y}| |\mathcal{S}|}{4n\mu_n^2}\right).
    \end{IEEEeqnarray}
    We then take the expectation with respect to $P_{\tilde{M}}$ and use the 
  convergence of $(\EE[\pi_{\tilde{X}^{n_i}}(x)])_{n_i \geq 1}$ to $P_X$ to conclude:
    \begin{equation}
 \varliminf_{i\to +\infty}    \dfrac{1}{n_i} H(\tilde{Y}^{n_i} |\tilde{M} )  \geq \sum_{a \in \mathcal{X}} H(Y|X = a) P_X(a). 
    \end{equation}

\end{document}